\documentclass[prb,showpacs,superscriptaddress,twocolumn,notitlepage]{revtex4-1}

\usepackage[english]{babel}
\usepackage[utf8x]{inputenc}
\usepackage[T1]{fontenc}

\usepackage{amsmath}
\usepackage{graphicx}

\usepackage{natbib}
\usepackage[final=true]{hyperref}

\begin{document}
%\maketitle

\title{Ultracentrifugation for ultrafine nanodiamond fractionation}

\author{S.~V.~Koniakhin}
\email{kon@mail.ioffe.ru}
\affiliation{St. Petersburg Academic University - Nanotechnology Research and Education Centre of the Russian Academy of Sciences, 194021 St. Petersburg, Russia}
\affiliation{Ioffe Physical-Technical Institute of the Russian Academy of Sciences, 194021 St.~Petersburg, Russia}
\affiliation{Institut Pascal, PHOTON-N2, University Clermont Auvergne, CNRS, 4 avenue Blaise Pascal, 63178 Aubi\`{e}re Cedex, France.} 

\author{N.~A.~Besedina}
\affiliation{St. Petersburg Academic University - Nanotechnology Research and Education Centre of the Russian Academy of Sciences, 194021 St. Petersburg, Russia}

\author{D.~A.~Kirilenko}
\affiliation{Ioffe Physical-Technical Institute of the Russian Academy of Sciences, 194021 St.~Petersburg, Russia}

\author{A.~V.~Shvidchenko}
\affiliation{Ioffe Physical-Technical Institute of the Russian Academy of Sciences, 194021 St.~Petersburg, Russia}

\author{E.~D.~Eidelman}
\affiliation{Ioffe Physical-Technical Institute of the Russian Academy of Sciences, 194021 St.~Petersburg, Russia}
\affiliation{St. Petersburg State Chemical Pharmaceutical Academy, 197376 St.~Petersburg, Russia}

\begin{abstract}
In this paper we propose a method for ultrafine fractionation of nanodiamonds using the differential centrifugation in the fields up to 215000g. The developed protocols yield 4-6 nm fraction giving main contribution to the light scattering intensity. The desired 4-6 nm fraction can be obtained from various types of initial nanodiamonds: three types of detonation nanodiamonds differing in purifying methods, laser synthesis nanodiamonds and nanodiamonds made by milling. The characterization of the obtained hydrosols was conducted with Dynamic Light Scattering, Zeta potential measurements, powder XRD and TEM. According to powder XRD and TEM data ultracentrifugation also leads to a further fractionation of the primary diamond nanocrystallites in the hydrosols from 4 to 2 nm.
\end{abstract}

%while previously only aggregates larger than 20 nm were dominant in optical experiments including dynamic light scattering. 

\maketitle

\section{Introduction}

Nanodiamonds are a promising type of ultrasmall nanoparticles and they are actively investigated for application in advanced materials fabrication \cite{mochalin2012properties}, quantum computing \cite{Chen2017}, biology and medicine\cite{ho2010applications}. Most number of nanodiamonds applications assumes their handling in the from of hydrosols. It is also well known that the particle size distributions in the detonation nanodiamond hydrosols and the hydrosols of the other nanodiamond types are typically strongly broadened. This picture stems from the fact that there are both individual primary crystallites of about 4-5 nm size \cite{shenderova2011nitrogen,osswald2009phonon,stehlik2015size,aleksenskii1997diamond,aleksenskii1999structure,vul2006direct,ozerin2008x,shenderova2015science,dideikin2017rehybridization,korepanov2017carbon} and their agglomerates of the size of 100 nm in the nanodiamond hydrosols\cite{koniakhin2015molecular,vul2011absorption}. The presence of  agglomerates significantly complicates the application of nanodiamonds and their characterization, especially with optical methods. According to the Rayleigh theory the scattering cross section is proportional to the square of the scattering particle volume \cite{hulst1957light,landau2013electrodynamics,bohren2008absorption}. Thus, despite the small total number of  agglomerates and their small weight fraction,  agglomerates produce the dominant contribution to the light scattering in the hydrosols of nanodiamonds. This can be seen from the analysis of static \cite{vul2011absorption,aleksenskii2012optical,konyakhin2013labeling} and dynamic light scattering (DLS) data \cite{koniakhin2015molecular,aleksenskii2012applicability,osawaDLS}. Obviously, presence of  agglomerates complicates the size determination of primary nanodiamond crystallites using the DLS method\cite{osawaDLS}.

These  agglomerates are very stable and cannot be broken even by ultrasound treatment \cite{shenderova2015science}, so chemical deagglomeration and annealing \cite{aleksenskiy2011deagglomeration,dideikin2017rehybridization} or bead-milling \cite{osawa2008monodisperse,kruger2005unusually} is required for their elimination. Even after such processes a significant amount of  agglomerates remains in nanodiamond powders and hydrosols. In the case of complete removal of the nanodiamond  agglomerates from the hydrosol, the signal from the primary nanodiamond crystallites can be detected by the DLS method.

\begin{figure}
\centering
\includegraphics[width=1.0\linewidth]{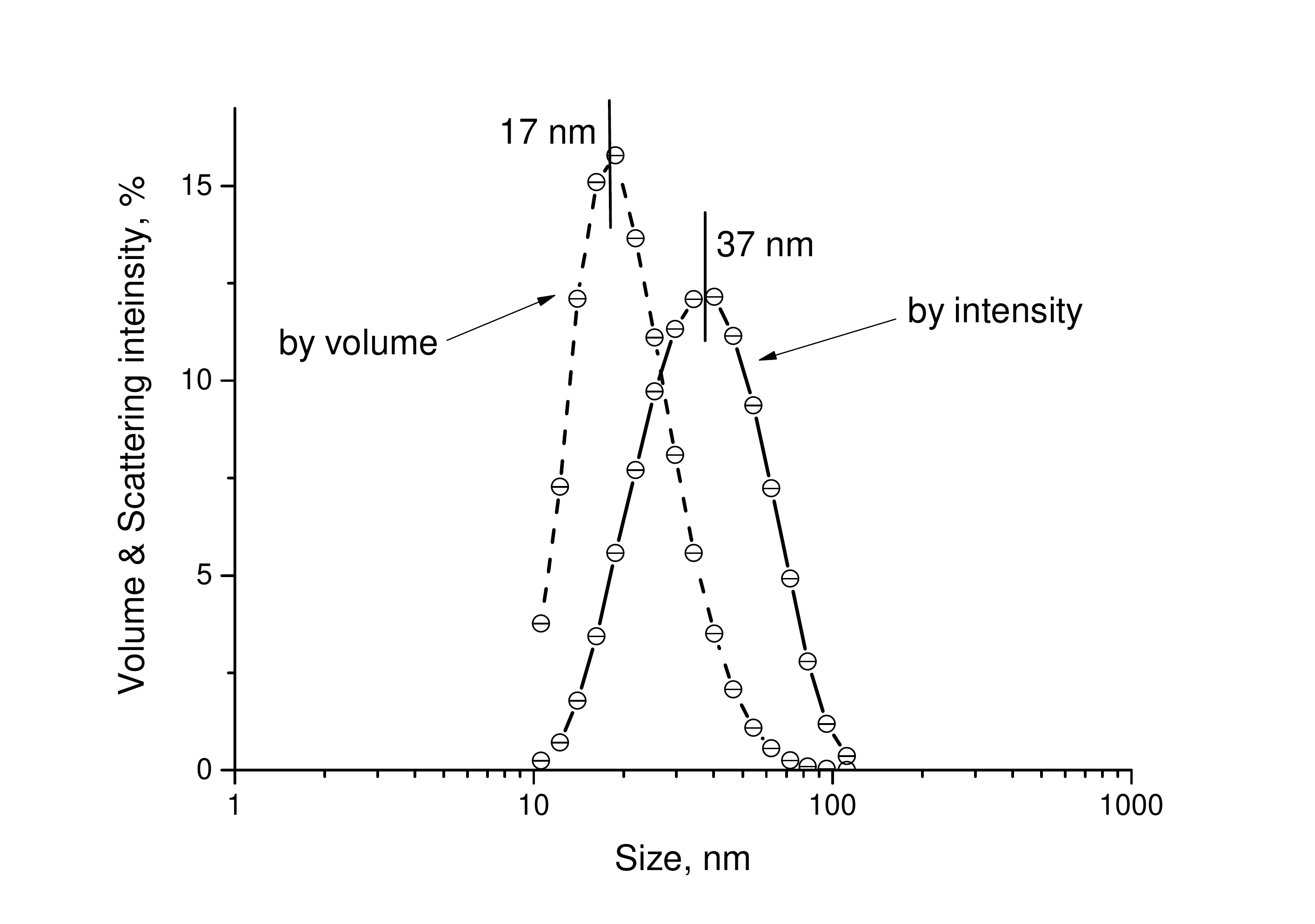}
\caption{\label{Fig1} Particle size distribution in initial Z- nanodiamond hydrosol in terms of light scattering intensity fraction and volume fraction obtained by DLS method. Volume distribution is shifted to the smaller size with respect to light scattering intensity distribution due to proportionality of scattering cross section to the square of nanoparticle volume.}
\end{figure}

In the literature, particle size distributions are usually given in terms of the volume fractions of $P_{vol}(D)$ of particles with size $D$. These distributions can be transformed into light scattering intensity distributions as $P_{int}(D) \propto P_{vol}(D) \cdot D^3$. After recalculation, the peak inevitably shifts to the region of larger particles. This is evident from Fig. \ref{Fig1}, which shows the particle distributions obtained by DLS method for one of the nanodiamond types considered in this paper (deagglomerated detonation nanodiamond with a negative $\zeta$-potential, Z-). The distributions are given by the volume fraction and by the fraction of the scattered light intensity. Even for suspensions, which contain mainly approx. 4 nm particles, the maximum of scattering intensity falls on the sizes larger than 10 nm. It is important to underscore that the lower-level result of DLS is the scattered light intensity distribution. The volume distributions are obtained using recalculation. Here, we use particle size distributions in terms of light scattering intensity to highlight the absence of scattering by agglomerates.

A significant progress in reducing nanodiamond particles size has been achieved with the use of annealing \cite{stehlik2015size,stehlik2016high}. However, the distributions given in these papers were obtained by the local method of atomic force microscopy \cite{stehlik2016size} and this data can not be directly related to DLS method results.

Previously, the centrifugation effectiveness for nanodiamond fractionation was shown in Refs. \cite{korobov2013improving,williams2010size}. The relative centrifugal force (RCF) up to 15000g was used and the usage of larger fields is of high interest.

The ultracentrifugation-based method for nanodiamond fractionation was developed by Peng \textit{et al.} in Ref.\cite{peng2013gram}. The authors proposed the effective two stage rate-zonal density gradient ultracentrifugation (RZDGU) protocol giving high yield after the first stage and moderate yield after the second one. The protocol required dialysis to remove the sucrose from the solvent.

In this paper we propose the protocols for the highly effective fractionation of various-typed nanodiamond hydrosols using ultracentrifugation in the fields up to 215000g and provide the characterization of the obtained material using DLS, $\zeta$-potential measurements, XRD diffraction, and TEM. We demonstrate a possibility to obtain the nanodiamond fraction with lower than 10 nm size giving main contribution to light scattering, which can be straightly detected by DLS method. The various nanodiamond types can be used for such procedure as initial samples. The developed protocols imply nanodiamond centrifugation in native solvents, which prevents contamination and excludes additional purification procedures like dialysis.

The obtained hydrosols are characterized by DLS as consisting of nanoparticles lower than 10 nm only, while for the hydrosols before ultracentrifugation maximum of light scattering intensity is at 40 nm and higher.

\section{Experimental}

\subsection{Samples}

In this paper, the hydrosols of various nanodiamond types were taken for ultracentrifugation to demonstrate a possibility of lower than 10 nm fraction separating. The following detonation nanodiamond types were investigated: chemically purified industrial detonation nanodiamond (Z0), the same diamond chemically purified and deagglomerated by annealing in Hydrogen (positive  $\zeta$-potential, Z+) and chemically purified and deagglomerated by annealing in air diamond (negative  $\zeta$-potential, Z-). The detailed information about Z0, Z+ and Z- samples is given in Refs. \cite{vul2017transition,dideikin2017rehybridization}. Also, nanodiamonds of laser synthesis (Laser) \cite{baidakova2013structure} and nanodiamonds made by milling (Milling) \cite{kruger2005unusually,osawa2008monodisperse} were studied.

\subsection{Ultracentrifugation}

Fractionation of the initial nanodiamond hydrosols was carried out using a preparative ultracentrifuge Beckman Coulter Optima L-100 XP equipped with a SW 60 Ti swinging bucket rotor designed for 6 tubes of 4 ml volume. Fractionation was conducted in the differential centrifugation mode in native solvent (deionized water). Under the action of effective gravity, the particles in the suspension acquire an almost constant sedimentation velocity and drift from the top of the tube to its bottom. The equilibrium of gravity and the hydrodynamic friction force (Stokes force) takes place \cite{kowalczyk2011nanoseparations}. The expression for the sedimentation (pelleting) time $T_{pell}$ for the particles suspended at a distance $r_{min}$ from the rotor rotation axis to the tube bottom at a distance $r_{max}$ reads as

\begin{equation}
T_{pell}=\frac{18\eta \ln (r_{max}/r_{min})}{(\rho_{D}-\rho_{w}) \omega^2 D^2},
\end{equation}
where $\omega$ is rotor angular velocity, $\eta$ is water viscosity, $D$ is nanoparticle hydrodynamic diameter, $\rho_D$ and $\rho_W$ are mass densities of bulk diamond and water, respectively.

After centrifugation, the tubes were removed from the rotor, and the upper 1 ml of supernatant was taken by a pipettor for characterization by dynamic light scattering and other methods. The used rotation speed and the centrifugation time lead to the solid pellets formation on the bottom of the tubes. These pellets can be partly redispersed in water only by ultrasonic treatment.

The first developed protocol (40k-3h) consisted in centrifugation of a homogeneous suspension at a speed of 40000 RPM for $T$ = 3h and taking the supernatant from the tube. This speed and the geometric dimensions of the rotor correspond to the maximum field at the bottom of the tube 215000g. These conditions correspond to a ratio of $T/T_{pell} = 0.72$ for 4 nm particles and $T/T_{pell} = 72$ for 40 nm  agglomerates. The sedimentation coefficient of 4 nm nanodiamonds is approximately 25 Svedberg.

The ratio of the actual centrifugation time $T$ to the $T_{pell}$ characterizes the hardness of the centrifugation conditions. At $T/T_{pell} = 1$, nearly all particles with a given size should settle at the bottom of the tube.

The purity of the final product is strongly influenced by mechanical manipulations with supernatant after centrifugation, such as extracting tubes from the rotor and taking the supernatant with a pipettor. These manipulations can cause the penetration of agglomerates from the precipitate into the supernatant. To prevent this effect, a homogeneous viscous solution of 0.5 ml of polyethylene glycol was added to the bottom of the tube, reducing the mobility of the settled particles and preventing their return to the supernatant. The weight ratio in a viscous solution of deionized water and dry PEG 8000 flakes was 1:1. The corresponding protocol for analogous rotor speed and centrifugation time is designated as 40k-3h-PEG.

Also, milder conditions were applied for centrifugation. In the 40k-1h protocol the rotor speed was also 40000 RPM and the centrifugation time was reduced to 1 hour. Protocol 40k-1h-PEG implies centrifugation with the same parameters and addition of a PEG layer to the tube bottom. The conditions in these protocols correspond to a ratio $T/T_{pell} = 0.24$ for 4 nm diamond particles and $T/T_{pell} =24$ for 40 nm. Finally, in protocols 23k-1h and 23k-1h-PEG, centrifugation at a rotor speed of 23000 RPM (71000 RCF) is carried out for one hour. In this case, the ratio $T/T_{pell} = 0.08$ for 4 nm particles and $T/T_{pell} = 8$ for 40 nm particles.

\subsection{Characterization techniques}

Characterization of samples before and after centrifugation was carried out using the following methods. The size distributions were obtained by dynamic light scattering method with the Malvern Zetasizer ZS 3600 instrument. As it was shown previously, the DLS method yields accurate estimates of the nanoparticles size down to 3 nm \cite{koniakhin2015molecular}. For smaller nanoparticles, the accuracy remains at the level of 20\%. Thus, the DLS method provides the sufficient accuracy to characterize the suspensions obtained by ultracentrifugation. For $\zeta$-potential measurements (using Malvern Zetasizer ZS 3600) 50 ml of Z+ 40K-3h and Z- 40K-3h samples were concentrated using a rotary evaporator to 2 ml with B\"uchi Rotavapor R-200.

The additional characterization by the means of XRD diffraction and TEM was provided for Z+ nanodiamond after ultracentrifugation using 40k-3h protocol. The Bruker Smart Apex Duo instrument equipped with Cu K-alpha source and a two-dimensional Apex detector was used. The experiment was performed in Debye-Scherrer geometry (transmission). The sample was fixed with a Zaponlack (nitrocellulose lacquer) at the end of the cactus needle. These measures minimize the contribution from extraneous ordered diffractive substances. One sees that the characteristic XRD fingerprints of the needles \cite{gindl2012structure} at 17, 23 and 34 degree do not interfere with the diamond diffraction pattern. For powder XRD studies the Z+ 40K-3h nanodiamond was dried in air. Transmission electron microscopy (TEM) studies were performed using Jeol JEM-2100F microscope (accelerating voltage 200 kV, point-to-point resolution 0.19 nm). Specimens for TEM were prepared by deposition of the nanodiamond hydrosol onto a conventional carbon lacey film.

\section{Results and discussion}

\subsection{DLS measurements}

In this section the DLS measurements of initial nanodiamond hydrosols and the hydrosols after ultracentrifugation are described.

\begin{figure}
\centering
\includegraphics[width=1.0\linewidth]{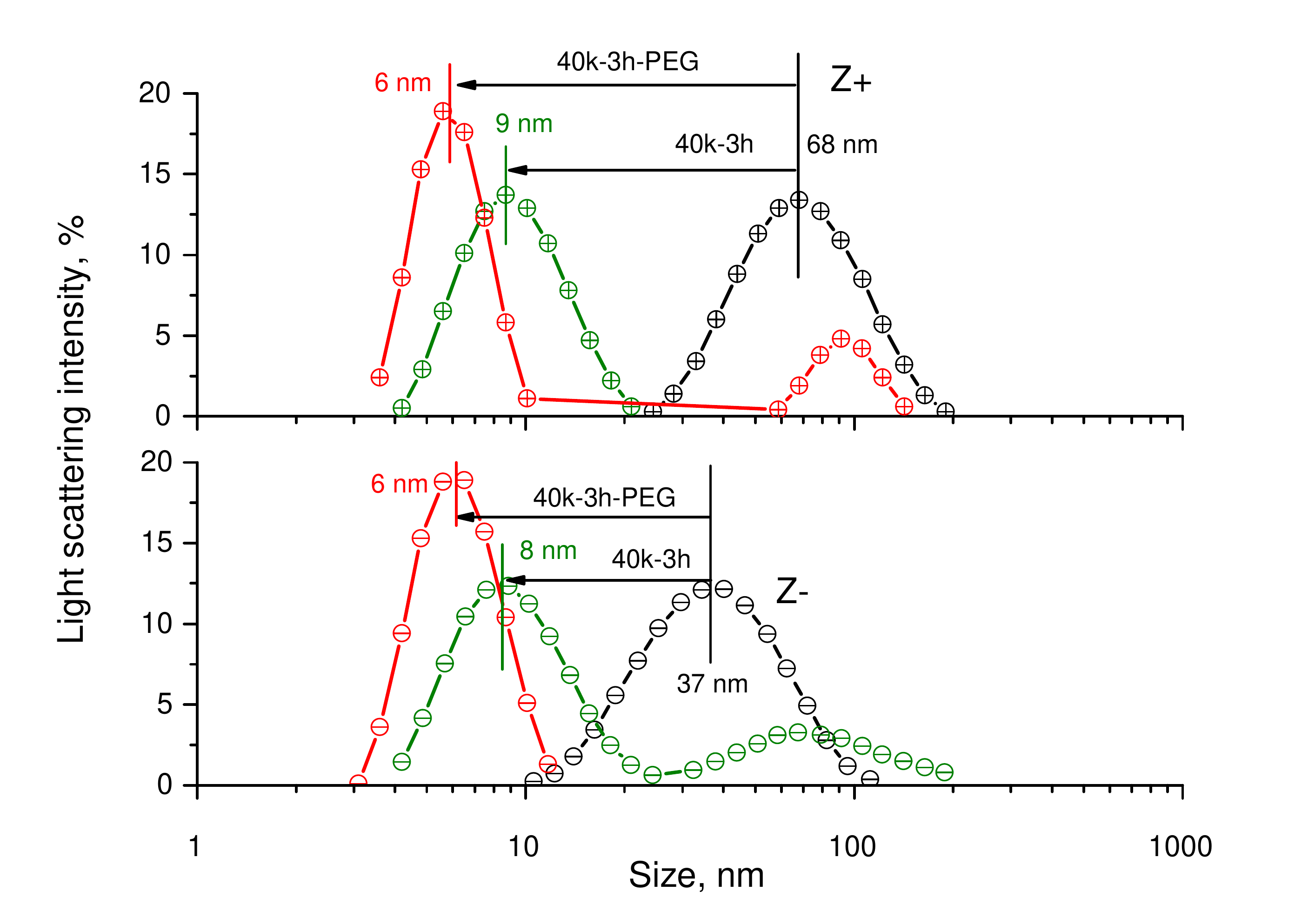}
\caption{\label{Fig2} Results of Z+ and Z- nanodiamond hydrosols ultracentrifugation using the 40k-3h and 40k-3h-PEG protocols (40000 RPM rotor speed, which corresponds to 215000g centrifugation field, 3 hours duration, without PEG and with viscous PEG layer on the tube bottom). The obtained with DLS distributions indicate that the desired fraction of the nanoparticles lower than 10 nm in size gives the main contribution to the light scattering intensity in the hydrosols.}
\end{figure}

\begin{figure}
\centering
\includegraphics[width=1.0\linewidth]{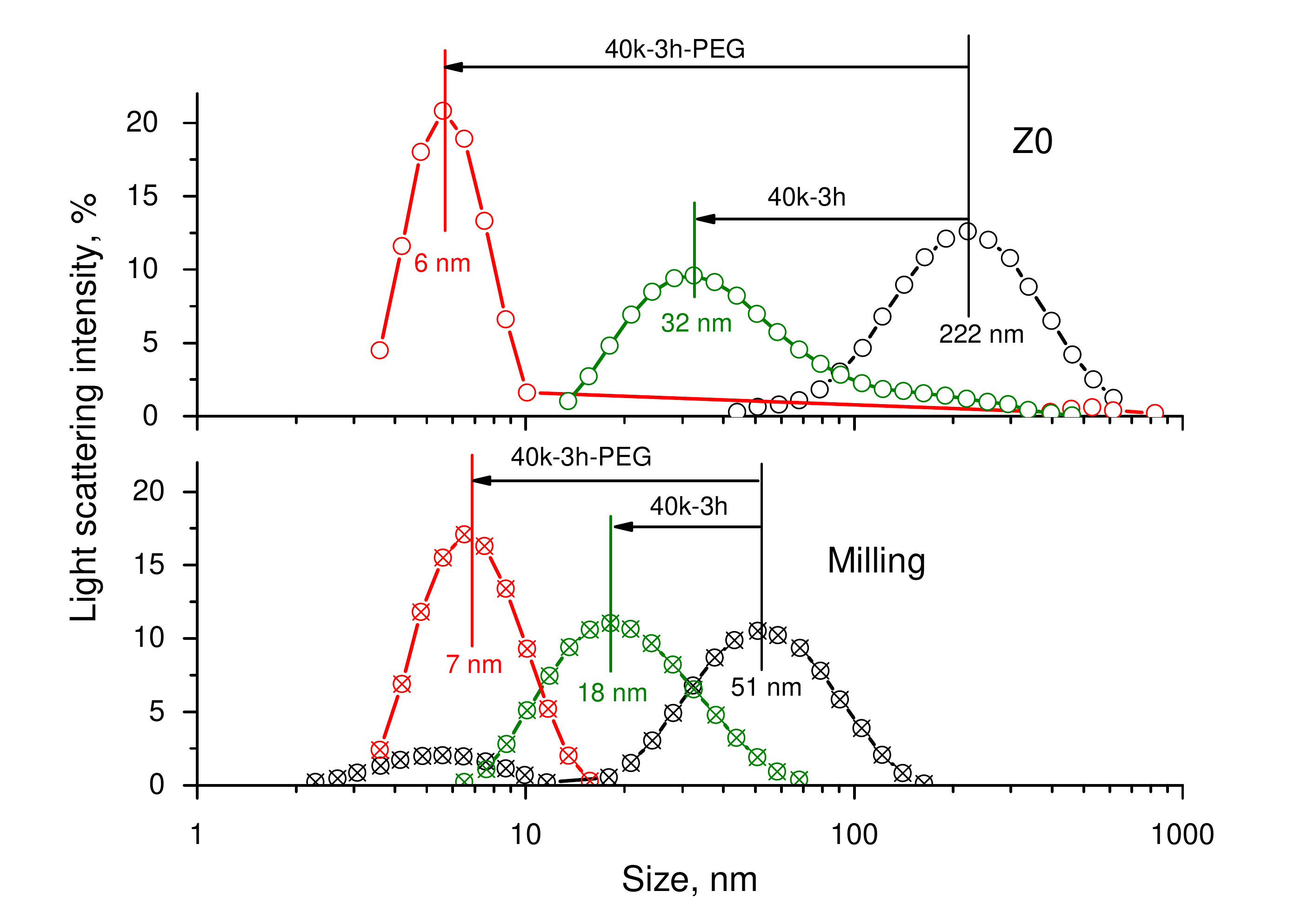}
\caption{\label{Fig3} Results of the Z0 nanodiamond hydrosol and the hydrosol of nanodiamonds made by milling ultracentrifugation using the 40k-3h and 40k-3h-PEG protocols. Adding the PEG layer onto the tube bottom is required for obtaining fraction lower than 10 nm.}
\end{figure}

\begin{figure}
\centering
\includegraphics[width=1.0\linewidth]{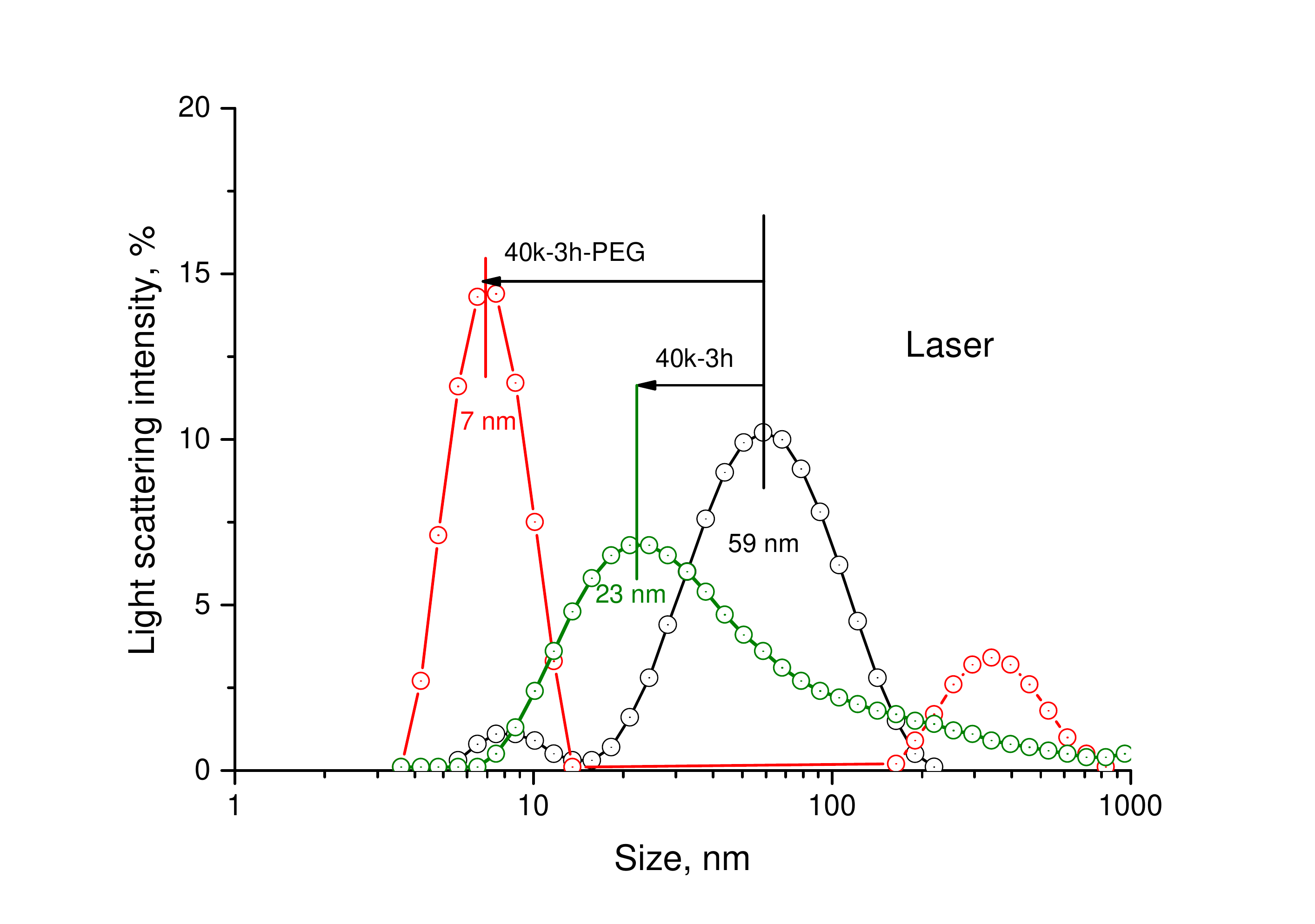}
\caption{\label{Fig4}  Results of the laser synthesis nanodiamond hydrosol ultracentrifugation using the 40k-3h and 40k-3h-PEG protocols.}
\end{figure}

Fig. \ref{Fig2} shows the scattered light intensity distributions by size for Z+ and Z- nanodiamonds ultracentrifuged using the 40k-3h and 40k-3h-PEG protocols. It can be seen that the fraction 6 nm in size gives the main contribution to scattering intensity was obtained with 40k-3h-PEG protocol. This size is very close to the size of an individual diamond nanocrystallite \cite{dideikin2017rehybridization}. Fig. \ref{Fig3} is similar to the previous one and shows the ultracentrifugation results for Z0 sample and for nanodiamonds made by milling. Finally, in Fig. \ref{Fig4} the results of laser synthesis diamond ultracentrifugation are shown.

Interestingly, the fraction smaller than 10 nm can be obtained from all the samples investigated regardless the initial sample size distribution, which is an evidence of approx 5 nm size of primary crystallites for various nanodiamond types.

\begin{figure}
\centering
\includegraphics[width=1.0\linewidth]{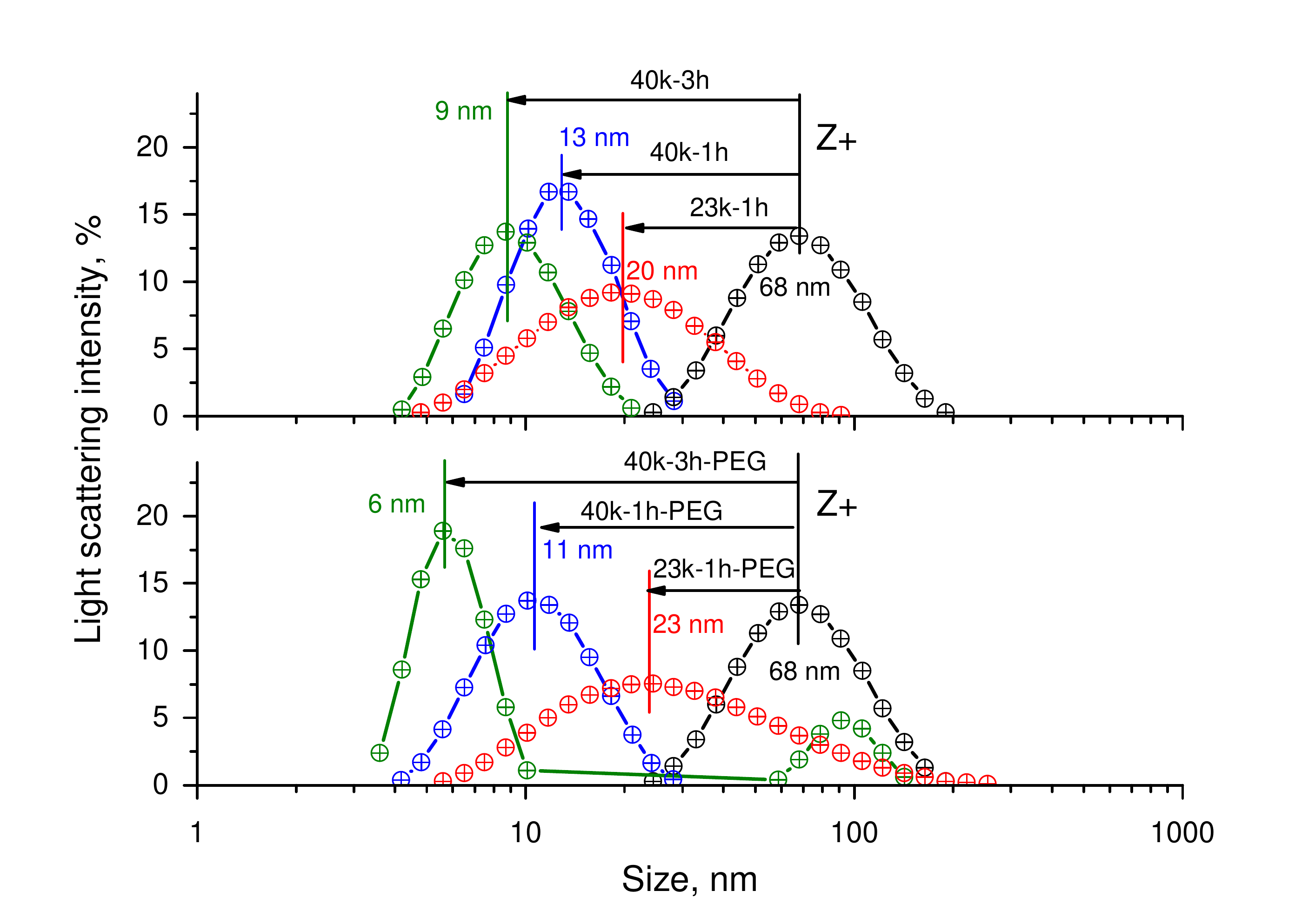}
\caption{\label{Fig5}  Results of Z+ nanodiamond hydrosol ultracentrifugation using protocols differing in duration time and rotor speed. Hardening the ultracentrifugation conditions leads to separating smaller fraction. Adding the PEG layer onto the tube bottom also leads to obtaining smaller fraction due to preventing the nanoparticles return to the supernatant from the pellet. The upper panel is for the protocols without PEG usage and the lower panel is for the protocols with PEG layer on the tube bottom.}
\end{figure}

It is important to analyze the effect of centrifugation conditions (rotor speed and duration time) and the presence of a viscous polyethylene glycol layer on the tubes bottom on the hydrosols fractionation results. Fig. \ref{Fig5} shows the results of Z+ nanodiamond hydrosol ultracentrifugation with the hardest employed protocol 40k-3h, softer protocols 40k-1h and 23k-1h and the relevant protocols with the addition of a viscous layer of polyethylene glycol (40k-3h-PEG, 40k-1h-PEG, 23k-1h-PEG). Application of polyethylene glycol viscous layer allows obtaining the hydrosols distributions corresponding to a smaller particle size. As can be seen, the use of polyethylene glycol gives more gain in obtaining smaller particles for more hard centrifugation conditions. When the conditions are softened, the influence of the viscous layer weakens.

\begin{figure}
\centering
\includegraphics[width=1.0\linewidth]{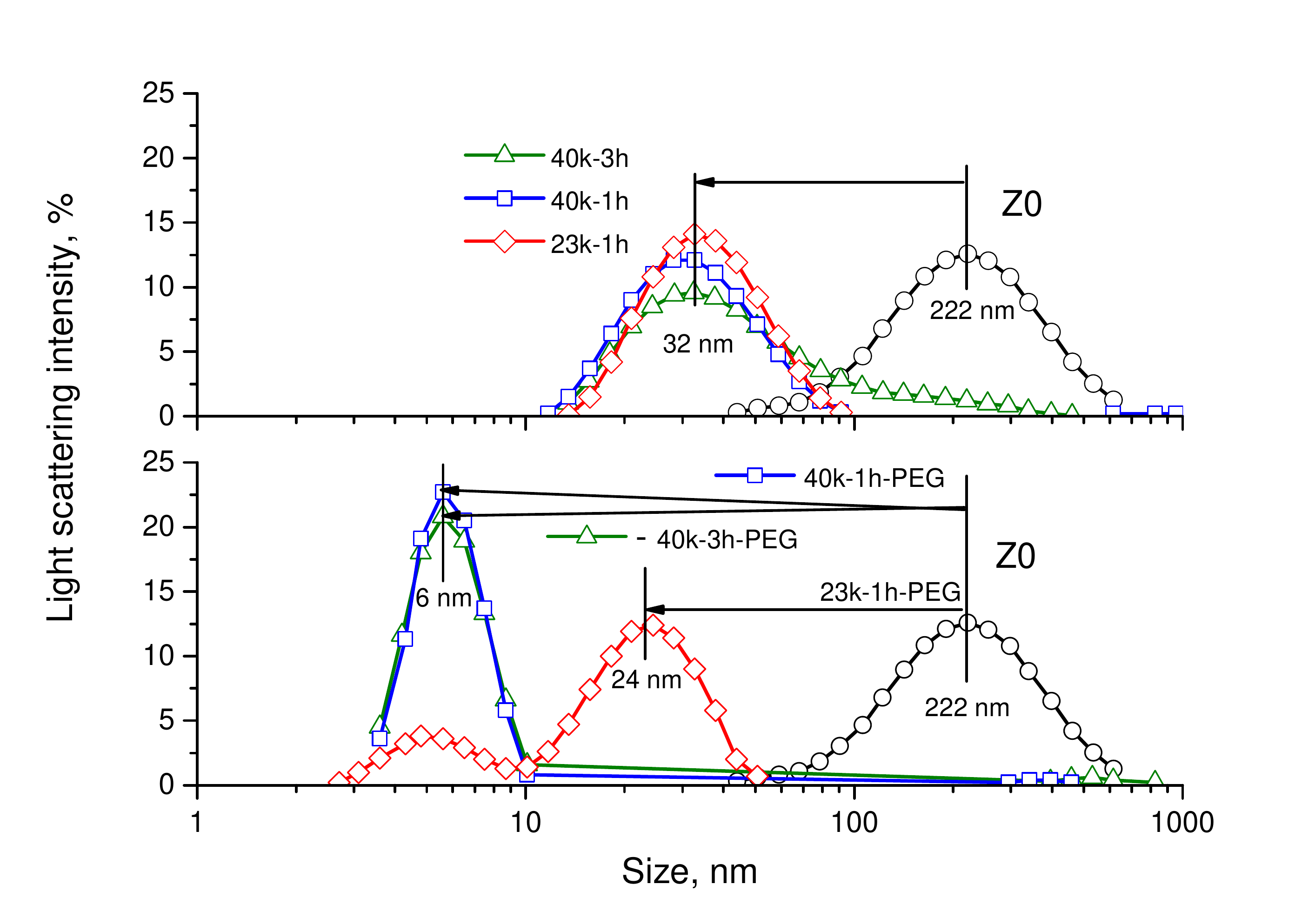}
\caption{\label{Fig6}  Results of Z0 nanodiamond hydrosol ultracentrifugation using protocols differing in duration time and rotor speed. The upper panel is for the protocols without PEG usage and the lower panel is for the protocols with PEG layer on the tube bottom.}
\end{figure}

The picture for the Z0 diamond hydrosol with the characteristic  agglomerate size more than 100 nm is principally different, see Fig. \ref{Fig6}. Centrifugation without PEG layer can not yield fraction smaller than 10 nm. On the contrary, even 40k-1h-PEG protocol successes in obtaining small fraction from Z0 nanodiamond.

Using the protocols with PEG layer yields smaller sized nanodiamonds in the supernatant, however the dialysis of supernatants for removing the dissolved PEG is strictly recommended, which complicates overall fractionation procedure. Therefore for the further characterization the samples obtained with 40k-3h protocols were used.

\subsection{Characterization of nanodiamonds after ultracentrifugation}

The samples obtained after centrifugation were very dilute and almost colorless. Concentration with rotary evaporator allowed 25-fold increasing the concentration and simplified further characterization including $\zeta$-potential measurements. After concentration, the samples gained the yellow-brown color. Noticeably, no aggregation during evaporation process occurred. Z+ 40k-3h and Z- 40k-3h samples revealed the Zeta potential values +31 mV and -46 mV, respectively. These values of $\zeta$-potential give the evidence of the produced hydrosols stability.

\begin{figure}
\centering
\includegraphics[width=1.0\linewidth]{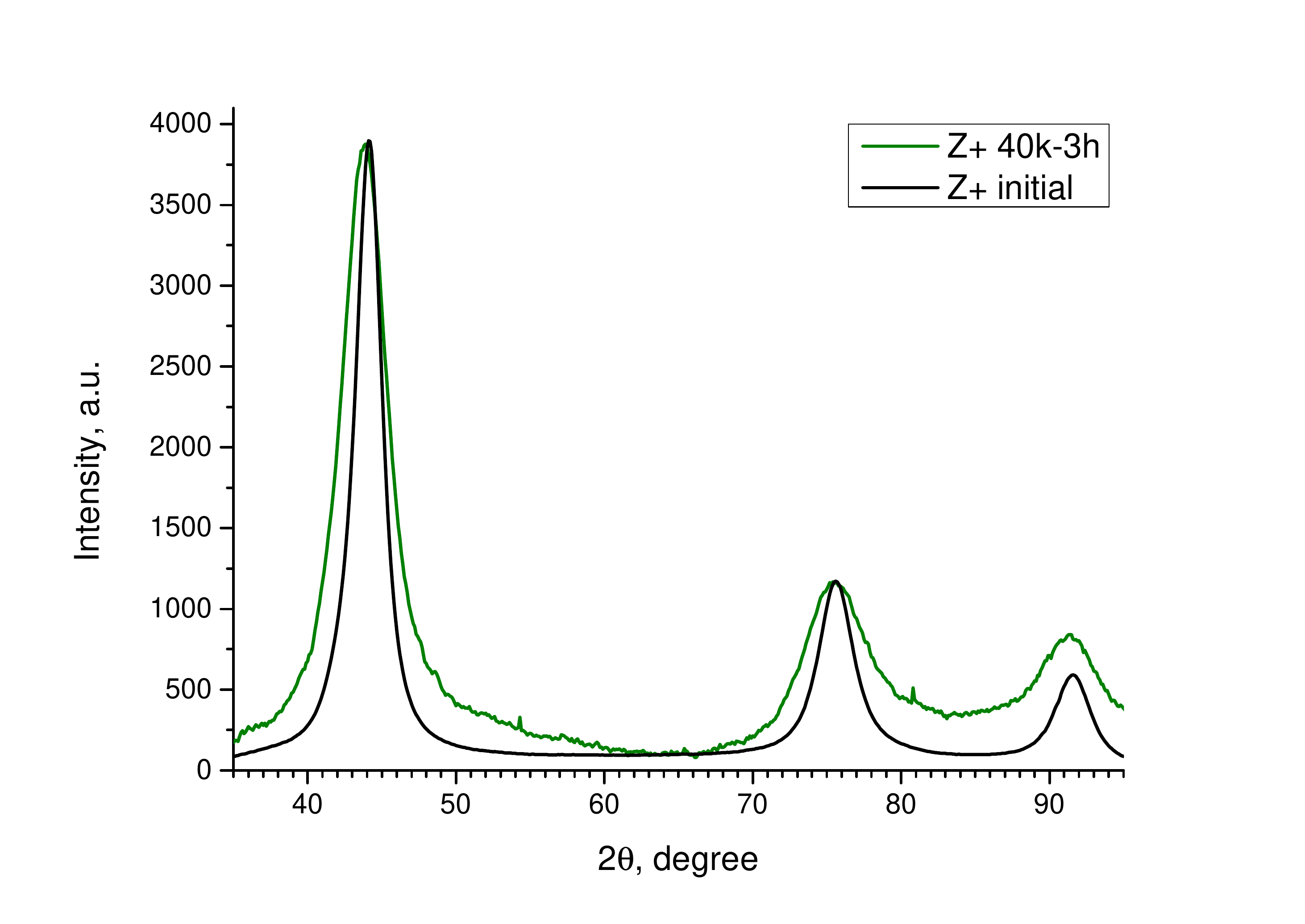}
\caption{\label{figXRD} XRD pattern of Z+ nanodiamond after 40k-3h protocol in comparison with initial Z+ nanodiamond. Sherrer equation yields 2.2 nm for Z+ 40k-3h nanodiamond size and 3.9 nm for initial Z+ nanodiamond. Ultracentrifugation provides the additional fractionation of the primary nanodiamond crystallites.}
\end{figure}

In Fig. \ref{figXRD} the data on powder diffraction of the initial Z+ sample and Z+ sample after ultracentrifugation using protocol 40k-3h is shown. The peak positions correspond to the diamond lattice (approx. 43.8$^{\circ}$, 75.7$^{\circ}$, 91.2$^{\circ}$). The initial Z+ nanodiamond (111) peak coincides with one given in Ref. \cite{stehlik2015size} (see Fig. S1 in supplementary of Ref. \cite{stehlik2015size}, sample DND-asrec) and with typical profiles in Ref. \cite{ozerin2008x}. Also it resembles the XRD data from Ref. \cite{baidakova2007new}. For Z+ after 40k-3h protocol the peaks reveal almost twofold broadening. The values of FWHM substituted to the Scherrer euqation indicate a characteristic coherent scattering region size of about 2.2 nm for diamond after ultracentrifugation and 3.9 nm for initial Z+ diamond. The data was averaged over three peaks. Thus, one sees that ultracentrifugation leads to a further fractionation of the primary nanocrystallites in hydrosols. 

The powder XRD data and the correlation between $\zeta$-potential of initial nanodiamonds Z+ and Z- and the nanodiamonds after ultacentrifugation justify that ultracentrufugation process preserves the nature of the diamond nanoparticles in hydsosols and no contamination occur.

\begin{figure}
\centering
\includegraphics[width=1.0\linewidth]{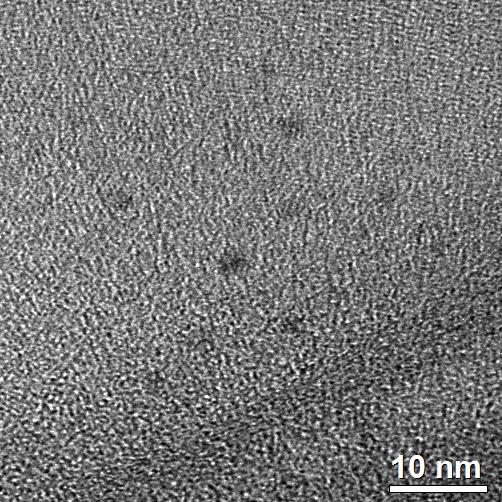}
\caption{\label{figTEM} TEM image of Z+ diamond after 40k-3h protocol.}
\end{figure}

Finally, Fig. \ref{figTEM} shows the TEM picture of Z+ diamond after ultracentrifugation using 40k-3h protocol. The observed particles size coincides with data extracted from the XRD data. 

\section{Conclusion}

Using ultracentrifugation, a fraction of diamond nanoparticles of less than 10 nm in size was obtained from all nanodiamond types used: chemically purified industrial detonation nanodiamonds, deagglomerated industrial detonation nanodiamonds with positive and negative $\zeta$-potential, nanodiamonds of laser synthesis and nanodiamonds made by milling. Thus we conclude the characteristic size of the primary individual nanocrystallites less than 10 nm occurs independently on the method for producing diamond nanoparticles. However the obtained sizes do not reach 4-5 nm and with high probability dimers, trimers and other nanodiamond oligomers remain in hydrosols.

A protocol consisting in 3 hours ultracentrifugation with RCF 215000 is required for the stable reproducible production of a practically monodisperse fraction of nanodiamonds of 6-9 nm size from the investigated samples. According to DLS data the obtained fraction gives major contribution to the light scattering, which is an evidence of complete vanishing of the agglomerates from the hydrosols. The protocols with addition of viscous PEG layer onto the bottom of the centrifugation tubes yield smaller nanodiamond fraction due to suppression the pellet distribution and particles diffusion and thus preventing the return of the large particles into supernatant. The significant advantage of the protocols without PEG is the absence of any contamination with PEG and the ability to provide centrifugation in any native solvent compatible with centrifuge rotor and tubes. 

Importantly, the fraction of 6 nm particles was obtained from Z0 sample by means of ultracentrufugation without any additional annealing as in methodology leading to Z+ nanodiamond (annealing in hydrogen) and to Z- nanodiamond (annealing in air), see Ref. \cite{dideikin2017rehybridization}. Therefore, ultracentrifugation is possibly an alternative to chemical deagglomeration and annealing for obtaining nanometer sized diamonds. However, it is important to underscore that Z+ and Z- nanodiamond hydrosols require the mildest centrifugation conditions including protocols without PEG to obtain the fraction less than 10 nm in size. It is the evidence of the high purification and deagglomeration quality of these nanodiamond types \cite{dideikin2017rehybridization}. Most likely, Z+ and Z- nanodiamond hydrosols contain the significant fraction of the free primary crystallites and small fraction of the agglomerates.

According to powder XRD and TEM data, the nanodiamonds size after ultracentrifugation reaches approx. 2 nm, which is an evidence of the additional ultrafine fractionation of the primary diamond nanocrystallites in hydrosols comparable to the results described in Ref.\cite{stehlik2016high}. This picture coincides with the hypothesis that approx. 3 nm is the stable size of diamond nanocluster\cite{raty2003ultradispersity}. The correlation between $\zeta$-potential of initial nanodiamonds and nanodiamonds after centrifugation and powder XRD data allow concluding that the nature of the diamond nanoparticles in hydsosols is preserved during the ultracentrifugation.

\begin{acknowledgments}

The authors acknowledge A.Ya. Vul for attention to work. We are gratefully indebted to O. Levinson and E. Osawa for providing the laser synthesis nanodiamonds and detonation nanodiamonds deagglomerated by milling method, respectively. The TEM studies have been carried out with the use of equipment of the Federal Joint Research Centre «Material science and characterization in advanced technology» (Ioffe Institute, St.-Petersburg, Russia). The authors acknowledge A.V. Nalitov, O.I. Utesov and A.E. Aleksenskii for fruitful discussions. Authors acknowledge the Russian Science Foundation (Project No. 16-19-00075). We are indebted to M.V. Dubina for support in providing experimental facilities.

\end{acknowledgments}

\bibliography{sample}

\end{document}